\documentstyle[12pt,cite]{article}

\newcommand{\be}{\begin{equation}}
\newcommand{\ee}{\end{equation}}
\newcommand{\bd}{\begin{description}}
\newcommand{\ed}{\end{description}}
\newcommand{\bea}{\begin{eqnarray}}
\newcommand{\eea}{\end{eqnarray}}
\newcommand{\ga}{\gamma}

\newcommand{\la}{\lambda}
\newcommand{\eps}{\epsilon}

\newcommand{\al}{\alpha}

\newcommand{\nn}{\nonumber}

\newcommand{\bna}{\begin{eqnarray*}}
\newcommand{\ena}{\end{eqnarray*}}

\date{}

\begin{document}
\thispagestyle{empty}
\begin{flushright}
ADP-95-41/T193 \\
hep-ph/9508074
\end{flushright}

\begin{center}
{\large{\bf Symbolic Algebra and Renormalization of Gauge Theories}} \\
\vspace{2.2 cm}
M. Rossi and A.P. Flitney \\
\vspace{1.2 cm}
{\it
Department of Physics and Mathematical Physics, University of Adelaide, \\
Adelaide, South Australia 5005, Australia \\}
\vspace{5mm}
\center{July 24, 1995}
\end{center}

\begin{abstract}
Symbolic algebra relevant to the renormalization of gauge
theories can be efficiently performed by machine using modern packages.
We devise a scheme for representing and
manipulating the objects involved in perturbative calculations of gauge
theories. This scheme is readily implemented using the general purpose
package, Mathematica. The techniques discussed are used to calculate
renormalization group functions for a non-abelian $SU(m)$ gauge theory
with massless fermions in a representation R, in the two-loop approximation,
and to simplify some expressions arising in electroweak calculations
at the two loop level.
\end{abstract}
\vspace{2.5cm}
\begin{flushleft}
Correspondence to A.P. Flitney at the address given \\
E-mail: {\it aflitney@physics.adelaide.edu.au}
\end{flushleft}

\pagebreak

\section{Introduction}

Perturbation theory has been an important calculational method in
quantum field theory for several decades. However the practical usefulness of
perturbation theory is limited mainly by the rapid rise in the amount of
labour required to improve the order of approximation. For instance, the
renormalization group functions can be calculated analytically to three
loops in non-abelian gauge theory, which requires the evaluation of 440 three
loop diagrams, and to five loops in $\la \phi^{4}$ theory~\cite{tar80,che81}.
The non-numerical nature of such perturbative calculations in quantum field
theory has complicated the effort to automate these calculations.
However, the development of specialised computer algebra systems~\cite{str74}
and improvements in general purpose computer algebra systems have greatly
facilitated this task~\cite{kub90,hy92,ber92,ft92,jam93,west93}.

In devising a scheme for the perturbative calculation of amplitudes in quantum
field theory the main areas which need to be developed are:
\begin{itemize}
\item The perturbation expansion itself,
\item Lorentz tensor, Dirac and symmetry group algebra, and
\item Feynman integration.
\end{itemize}
The perturbation expansion is either a Dyson-Wick expansion in the canonical
formalism or equivalently an expansion of the generating functional in
the path integral formalism. The perturbation expansion will not concern us
here, nor will we be concerned with automating symmetry group algebra which
is simple for all cases of interest to us. Our focus will be on Lorentz
tensor and Dirac algebra, and on Feynman integration.

There are a number of existing packages that tackle
the problem of Dirac algebra computations~\cite{kub90,hy92,ber92,jam93}.
The symbolic evaluation of Feynman diagrams at
tree level~\cite{ber92} and one loop~\cite{west93} have been
approached using Wolfram's computer algebra package Mathematica~\cite{wolf88}.
Fleischer and Tarasov~\cite{ft92} present a package for the evaluation of
certain two loop Feynman integrals written in the computer algebra language
FORM~\cite{ver90},
while symbolic three loop calculations~\cite{lar93}
have also been carried using FORM.
The current paper presents a package that combines a method for simplifying
the Dirac algebra with procedures for evaluating massless scalar or tensor
Feynman integrals at one or two loops,
and is implemented in Mathematica.

The following sections discuss our scheme for representing Feynman diagrams.
The scheme is essentially a definition of a notation that can be readily
expressed in Mathematica.
Section two introduces the notational scheme,
section three will discuss some aspects
of rules which have been encoded to simplify and integrate expressions,
and section four demonstrates the evaluation of some specific diagrams.
Usage messages for functions defined by us can be found in the appendix.

\section{Notational Scheme}

In this section it will be shown how one can use the symbolic
algebra capabilities of Mathematica to represent and simplify the
various types of tensorial and matrix expressions that arise in
amplitudes of gauge theories. This is based upon a suitable notation for
dimensionally continued tensors and Dirac gamma matrices. The aim is to
establish a notation that can be readily accommodated by the symbolic
capabilities of Mathematica. Dirac algebra can be performed without
reference to any representation of the Dirac matrices, by using the
commutation relations which define the algebra and identities derived
from those relations, along with the rules of matrix multiplication.
Tensor manipulations are also readily implemented symbolically. As we are
interested in evaluating amputated one particle irreducible amplitudes
which have been regularised using dimensional regularisation~\cite{tv72},
no explicit representation of Dirac gamma matrices is necessary.

The first step is to separate symbols which will represent
four-vectors from symbols which will represent tensor indices. We
generally use one or more lower case letters and numerals which are not
otherwise defined, such as {\tt a,k,p1 \ldots }  for four-vectors and indices.
The two types of symbols can be distinguished by declaring a list of symbols
which will represent four-vectors. Any symbols which
appear that are not in this list and are not otherwise defined shall
be understood to represent a Lorentz index, and need not be declared.
These symbols will appear only as the arguments of Mathematica
expressions. The expressions
themselves will represent Dirac matrices, Lorentz scalars,
four-vectors and tensors depending on the symbols that appear in their
arguments. We give the list of declared four vectors the name {\tt momenta}.

To illustrate this scheme assume {\tt momenta = \{k,p,q\} } throughout
the rest of this section. This list may be enlarged
at any time provided symbols that have already been used as Lorentz
indices are not included. Let the function {\tt g,} a
symmetric function of two arguments, denote the object that will
carry the properties of the metric tensor. What {\tt g} actually represents
will depend on its arguments as follows:
\begin{enumerate}
\item {\tt g[a,b]} represents the metric tensor with Lorentz
indices {\tt a} and {\tt b,} because {\tt a} and {\tt b} are not in the list
{\tt \{k,p,q\},}
\item {\tt g[a,k]} represents the four-vector {\tt k} with Lorentz
index {\tt a,} because only {\tt k} is in the list {\tt \{k,p,q\},} and
\item {\tt g[k,p]} represents the scalar product of the
four-vectors {\tt k} and {\tt p} which are both in {\tt \{k,p,q\}.}
\end{enumerate}

Products of gamma matrices are represented by a function
{\tt d} of any non-zero number of arguments. Again, what {\tt d} represents
depends
on the arguments. Wherever there appears an argument that is a member of
{\tt momenta} the corresponding matrix is the contraction of the gamma
matrix in that position with the four-vector. Otherwise the matrix is simply a
gamma matrix with the symbol denoting a Lorentz index. For example,
the object {\tt d[a,b,k,p,b,d,k,q]} denotes
\be
\gamma^{\alpha} \gamma^{\beta} \! \not \! k \not \! p \:
\gamma_{\beta} \: \gamma^{\delta} \! \not \! k \not \! q .
\ee
Free Lorentz indices appearing as arguments of {\tt g} and {\tt d}
can represent
contravariant or covariant indices. In renormalizing quantum field theories
we consider only amputated amplitudes. Information about components of
four-vectors and matrices is not required.
Thus the distinction between contravariant and
covariant components is not important here. One simply takes
the rank structure of the final expression to be that of the
input expression. If one wishes to consider expressions which included
external lines then one would need to distinguish between contravariant
and covariant indices. This could be done by declaring all covariant
indices in a list which one may call {\tt covariant.} Then symbols which
appear in neither of {\tt momenta} or {\tt covariant} and were not
otherwise defined would implicitly represent contravariant indices.

Traces of products of gamma matrices can be denoted by a function {\tt trace}
whose argument is a linear combination of products of gamma matrices.
This function will automatically convert the trace of a linear combination
to the linear combination of traces before any traces are evaluated.

Other tensors may be represented by other functions with the rank
structure given implicitly by the arguments. For example one may wish
to manipulate the Levi-Civita tensor in four dimensions. One would
define a function {\tt e} depending on four arguments.
Then, for example {\tt e[a,b,u,v]} would represent
\be
\epsilon^{\alpha \beta \mu \nu}
\ee
and {\tt e[a,b,k,p]} would represent
\be
\epsilon^{\alpha \beta \mu \nu} k_{\mu} p_{\nu}.
\ee

The mass of a particle is represented by the function {\tt M[x]}
where {\tt x} is a label for the convenience of the user
and is not used by the program.
Similarly, other scalars are represented by the function {\tt scalar[x].}
The left and right helicity projection operators are represented
by {\tt L} and {\tt R,} respectively. The matrix
$\gamma_{5}$ has not been explicitly represented,
nor have traces involving $\gamma_{5}$ been implemented. However the
definitions could be easily extended if traces involving $\gamma_{5}$ were
required (see for example~\cite{west93}).

Manipulation and simplification of expressions can be performed by
machine using pattern recognition, procedural programming, rule based
programming and functional programming all of which are supported by
Mathematica.
One can perform basic manipulations such as
\begin{itemize}
\item contracting repeated indices in tensor expressions,
\item simplifying an expression of the form
$ \gamma^{\mu} \gamma^{\alpha_{1}} \ldots \gamma^{\alpha_{j}}
\gamma_{\mu} $
\item commuting a gamma matrix through one or more gamma matrices,
\item the evaluation of the trace of any number of gamma matrices, and
\item evaluation of Feynman integrals
\end{itemize}
and any other lengthy algebraic manipulation that would be
prohibitively tedious to perform manually.  Some of the rule definitions
which are useful for the manipulations described are
illustrated in the next section. Note that the system function {\tt Dot}
has been used in place of {\tt d} for products of gamma matrices. This
function will perform matrix multiplication when explicit matrices in
component form are placed in its arguments. Since we require no
representation for the gamma matrices and thus only symbols appear as
its arguments, the function {\tt Dot} will simply represent non-commutative
multiplication and one can assign rules required to invoke the
various properties of Dirac algebra which are required. We have chosen to
use {\tt Dot} for this purpose because a convenient input notation
\be
{\tt Dot[a1, a2, \ldots] \; \equiv \; a1.a2. \, \ldots}
\ee
is available. Since {\tt Dot} has the attribute {\tt OneIdentity} the
expression
{\tt Dot[a]} is equivalent to {\tt a} where {\tt Head[a] = Symbol.} Hence a
solitary
symbol which is otherwise undefined represents a single gamma matrix.

Together with  symbols {\tt i} for the imaginary unit, {\tt n} for the
number of space-time dimensions, and {\tt eps = 4-n,} this scheme is
sufficient to represent any amputated diagram in a gauge theory. In
the next section we will discuss how to manipulate, simplify, and
integrate expressions within this scheme.

\section{Implementation of Symbolic Algebra}

We will begin by discussing expressions representing Lorentz tensors.
This will be followed with Dirac algebra and expressions involving
both gamma matrices and tensors, and traces of gamma matrices.
Finally the evaluation of two-loop Feynman integrals will be discussed.
In evaluating Lagrangian counterterms, which in turn give renormalization
group functions, we are interested in the pole part of amputated
diagrams. This necessitates the evaluation of Laurent expansions
and some points relating to this will be discussed briefly.

The metric of $n$-dimensional Minkowski space is represented by the
symmetric function {\tt g} with two arguments,
\bea
\lefteqn{{\tt Attributes[g] \,=\, \{Orderless\} }} & & \\
\lefteqn{{\tt g[x\_,x\_] \, := \, n}}  & {\tt /; \,FreeQ[momenta,x]} & \\
\lefteqn{{\tt g[a\_,b\_]\hat{\:}2 \, := \, n}} & & \nn \\
& \hspace{1.4cm}{\tt /; FreeQ[momenta,a] \: \&\& \: FreeQ[momenta,b]} & \\
\lefteqn{{\tt g[a\_,b\_]\hat{\:}2 \, := \, g[b,b]}} & & \nn \\
& \hspace{1.4cm}{\tt /; FreeQ[momenta,a] \: \&\& \: MemberQ[momenta,b] } &
\hspace{2.0cm}
\eea
Note that the action of the rules is conditional upon whether or not one or
both arguments are in the list {\tt momenta.} In conventional notation the
rules stated are
\bea
&& g^{\al}_{\al} = n  \\
&& g^{\al \beta} g_{\al \beta} = n  \\
&& b^{\al}b_{\al} = b^{2}
\eea
where b is a vector.  Rules can be applied at the discretion of the
user by defining new functions which act on tensors.  For example
the relations
\bea
&& g^{\al \beta} g_{\beta \la} = \delta^{\al}_{\la} \label{eq:rmind} \\
&& b^{\al} g_{\al \la} = b_{\la} \label{eq:lower}
\eea
can be implemented by
\bea
\lefteqn{ {\tt cc[x\_,a\_,b\_,s\_] \: := \: x \, /. \,g[a,s]\, g[s,b]\:
\rightarrow \: g[a,b]} } \nn \\
&&  \hspace{6.0cm} /; \, {\tt FreeQ[momenta,s]} \hspace{3.0cm}
\eea
Note that in this notation both (\ref{eq:rmind}) and (\ref{eq:lower}) are
embodied in one rule.
Functions can appear in the definitions of functions.  For instance, a
function which contracts all occurrences of a repeated index is
valuable. Suppose all the arguments of {\tt g} in an expression {\tt x}
are placed in a list called {\tt args} except for one repeated index {\tt s.}
We have defined a function, {\tt contract[x, args, s],} which
will perform all contractions of the repeated index {\tt s} in {\tt x.}
The definition of contract involves two other functions, {\tt cc} which
controls the actual replacement that is made, and {\tt ll} which takes a list
of symbols and returns a list of all pairs of those symbols. Extending
this further, we have defined a function,
{\tt contractall[x, repeated, other],} which takes an expression {\tt x,} a
list of all repeated indices placed in the second argument, and a list of
all remaining arguments of {\tt g} in {\tt x} in the third argument and
performs the contraction of all repeated indices. The
definition of this function is in terms of {\tt contract.}

As for {\tt g}, rules are assigned to {\tt Dot} to automatically implement
properties of matrix algebra. Note that since {\tt Dot} is a system function
it must be unprotected before rules can be added to its definition.
The rules we define for {\tt Dot} are applied automatically to any input
expression until no further changes occur.
These rules correspond to properties such as the distributive law over
addition and scalar multiplication. We have chosen to use the symbol
{\tt J} to denote the identity matrix.
Expressions which represent a combination of tensors and gamma matrices will
involve both {\tt g} and {\tt Dot.} In cases where a Lorentz index appears in
an argument of both {\tt g} and {\tt Dot} the index may be contracted out of
the
expression. A function {\tt slash} is defined in the appendix which performs
this task.

Identities, such as commutation relations can be applied at the
discretion of the user by defining functions which take an expression,
search for a specified pattern and replace the pattern by an equivalent
expression. For example, the commutation relations
\bea
&& \ga_{\al} \ga_{\beta} = - \ga_{\beta} \ga_{\al} + 2 g_{\al \beta}  \\
&& \ga_{\al} \ga_{\beta} \ga_{\la} = \ga_{\beta} \ga_{\la} \ga_{\al} +
2 \ga_{\la} g_{\al \beta} - 2 \ga_{\beta} g_{\al \la}
\eea
are applied by the function {\tt comm}, defined as
\bea
\lefteqn{ {\tt comm[x\_,a\_,b\_] \: := \: x \, /. \,a.b \rightarrow
 b.a \: + \: 2J \, g[a,b]}} & & \\
\lefteqn{{\tt comm[x\_,a\_,b\_,c\_] \: := \: x \, /. \, a.b.c \rightarrow
b.c.a \: + \: 2c\, g[a,b] \: - \: 2b\, g[a,c]}} & & \hspace{12.0cm}
\eea
Which of the two rules is applied depends on how many arguments are
passed to {\tt comm}. Note that the rules are valid whether or not
any of {\tt a, b} and {\tt c} appear in the list {\tt momenta.} Further rules
have been defined for commuting through more than two gamma matrices, and
another function called {\tt rcomm} for commuting in the reverse order.
Rules which apply identities for expressions of the form
\bea
&&\ga_{\mu} \ga_{\al_{1}} \ldots \ga_{\al_{j}} \ga^{\mu}  \\
&& k\!\!\!/ \ga_{\al_{1}} \ldots \ga_{\al_{j}} k\!\!\!/ \; , \;j=1,2,3,\ldots
\eea
have been similarly defined. For instance
\bea
{\tt con[x\_,a\_] \: := } & {\tt If \left[ \:MemberQ[momenta,a] \, ,
\right. }\hspace{5.5cm} &  \nn \\
& {\tt x\, /. \, a.a \rightarrow g[a,a]\, J,}\hspace{2.0cm} & \nn \\
& {\tt x\, /. \, a.a \rightarrow n\, J }\hspace{3.0cm} & \nn \\
& \lefteqn{\left. \right]} \hspace{8.5cm} &
\eea
\bea
{\tt con[x\_,a\_,b\_]\: := } & {\tt If \left[ MemberQ[momenta,a] \, ,
\right. } \hspace{5.5cm} & \nn \\
& {\tt x\, /. \, a.b.a \rightarrow  - b\, g[a,a]\: +\: 2a\,
g[a,b],} & \nonumber \\
& {\tt x\, /. \, a.b.a \rightarrow (2-n)\, b}\hspace{2.5cm} & \nn \\
& \lefteqn{\left. \right]} \hspace{8.5cm} &
\eea
The answer is dependent on whether or not the repeated symbol {\tt a}
appears in the list {\tt momenta.} Rules for $j>1$ can be included as needed.

To simplify expressions involving helicity projection operators,
the functions {\tt movel} and {\tt mover} have been defined.
They, respectively, move the helicity projection operators to the left or
to the right in each term in an expression.

The trace of an arbitrary linear combination of products of gamma matrices
can be evaluated using the linearity property of traces and the
recursive relation
\bea
Tr(\gamma^{\alpha_{1}} \gamma^{\alpha_{2}}) & = &
4 g^{\alpha_{1} \alpha_{2}}\\
Tr(\gamma^{\alpha_{1}} \ldots \gamma^{\alpha_{2j}}) & = & \sum_{l=o}^{2j-2}
(-1)^{l} g^{\alpha_{1} \alpha_{2j-l}} Tr(\gamma^{\alpha_{2}} \ldots
\hat{\gamma}^{\alpha_{2j-l}} \ldots \gamma^{\alpha_{2j}}) \label{eq:tr} \\
Tr(\gamma^{\alpha_{1}} \ldots \gamma^{\alpha_{2j+1}}) & = & 0
\eea
where the hat over the gamma matrix denotes its absence~\cite{west93}.
The function
{\tt trace} will firstly convert the trace of a linear combination into a
linear combination of traces, and then applies a function {\tt tr}
which makes use of the recursion relation (\ref{eq:tr}) to evaluate
the traces.
Note that {\tt tr} first checks that the length of the product is even and
returns zero if it is not. The case of a product of two gamma matrices
acts as an initial condition.

The scheme described so far can be used to represent any Feynman integral
which can arise in a gauge theory. A means of evaluating the integrals
would complete a scheme for the evaluation of amputated Feynman diagrams.
We will present a simple procedure which can be adapted to all cases.

We begin by considering integrals of the form
\be
\mu^{\la} \int \frac{d^{n}q}{(2 \pi)^{n}}
\frac{q_{\al} q_{\beta} \ldots}{(q^{2})^{j} ((q \pm p)^{2})^{l}}
\label{eq:stdint1}
\ee
and
\be
\mu^{\la} \int \frac{d^{n}q}{(2 \pi)^{n}}
\frac{q \!\!\!/ q_{\al} q_{\beta} \ldots}{(q^{2})^{j} ((q \pm p)^{2})^{l}}
\label{eq:stdint2}
\ee
where $\mu$ is an arbitrary mass scale, $q$ is the integration variable and
the rank of the tensor in the numerator is usually less than or
equal to 3. The factor $\mu^{\la}$ is
introduced to ensure that the integral is dimensionless for any $n$ (hence
$\la$ depends on $n$). This is sufficient for most
two loop calculations but if rules for the integration of higher rank
expressions are needed, then extra rules can be included.

The main consideration in defining an integration function is to be certain
that there is no residual dependence on the integration variable in the final
expression. For example, a function could be defined which searches for the
pattern
\be
{\tt \frac{1}{g[q,q] \: g[q+p,q+p]}}
\ee
and replaces it using a replacement rule. If that function were then
applied to a pattern like
\be
{\tt \frac{T[q]}{g[q,q] \:g[q+p,q+p]}}
\ee
the integration function would treat this as a product of two factors
\be
{\tt T[q]\; \frac{1}{g[q,q] \:g[q+p,q+p]}}
\ee
and replace only the second factor, leaving {\tt T[q]} in the resulting
expression. Hence it is important to check the final expression for the
presence of the integration variable.

In the integrals (\ref{eq:stdint1}) and (\ref{eq:stdint2}) there are other
considerations. The rank, the symbols used for indices, the integration
variable, the parameter {\tt p,} and the powers {\tt j} and {\tt l} should
all be variables which can be passed to the integrating function. We have
defined functions {\tt inttensor} and {\tt intgamma} which will return the
integrals of (\ref{eq:stdint1}) and (\ref{eq:stdint2}) respectively.
The arguments to be passed to {\tt inttensor} and {\tt intgamma} are
\be
{\tt inttensor[x, q, p, a, b, \ldots, j, l] }
\ee
and
\be
{\tt intgamma[x, q, p, a, b, \ldots, j, l], }
\ee
where {\tt x} is the expression to be integrated, {\tt q} is the integration
variable, {\tt p} is the four-momentum parameter, {\tt a,b,\ldots} are
Lorentz indices, and {\tt j} and {\tt l} are the powers in the denominator.
Again the definitions of {\tt inttensor} and {\tt intgamma} consist of a
limited series of rules. Which of the rules is applied depends on the number of
arguments passed to {\tt inttensor} and {\tt intgamma.} If the rules do not
cover cases of high rank that may be needed then new rules may be included
as needed. Rank three integrals are usually sufficient for most two-loop
calculations. Integrals involving non-zero masses have not been implemented,
though a function for these could be defined.

Dimensionally regularised integrals can be expanded as a Laurent series in
$\eps \; = \; 4-n$. In general an integral over $l$ four vectors will have a
leading pole of order $\eps^{-l}$. To evaluate counterterms only pole parts of
the Laurent expansions of integrals need be evaluated.
A table of pole parts of integrals can be prepared with the help of
{\tt inttensor} and {\tt intgamma.} This table can then be used to evaluate
the pole parts of diagrams and hence the counterterms. The integrals
required for this can be classified according to the form of the
denominators of the Feynman integrals. For example, to
calculate the two-loop boson self energy in a $SU(m)$ gauge
theory with massless fermions there is just one class of integrals
as the Feynman integral is always of the form
\be
\mu^{\la}
\int_{q,k} \frac{T_{\mu \nu}(q,k,p)}{q^{2}(q+p)^{2}k^{2}(k+p)^{2}(k-q)^{2}}
\label{eq;twopt}
\ee
where $p$ is the external momentum, and $T_{\mu \nu}(q,k,p)$ is a rank-2,
dimension 4 tensor constructed from $q$, $k$, and $p$. There are 75
distinct possibilities for $T_{\mu \nu}$ but fortunately many of the
corresponding integrals are related, and the number of independent integrals
is reduced to about 10. If we are interested only in the pole parts then
we can express any integral of the form (\ref{eq;twopt}) in terms of the pole
parts of just four integrals. We choose the four integrals where
$T_{\mu \nu} (q,k,p)$ is either
\be
t_{\mu \nu}(p) (q+p)^{2}
\ee
or
\be
t_{\mu \nu}(p) (k-q)^{2}
\ee
and $t_{\mu \nu}(p)$ is either $p_{\mu} p_{\nu}$ or $g_{\mu \nu} p^{2}$.

In the integrals discussed we have represented the Euler gamma function
by {\tt gam[x].} For the evaluation of pole parts of one-particle irreducible
amplitudes one must be able to perform the Laurent expansions. This can be
done by either replacing {\tt gam} with the system defined version of the
Euler gamma function and using such functions as {\tt Series,} or by defining
a sufficient set of rules for {\tt gam} to evaluate the expansion.

\section{Examples}

\begin{figure}

\begin{picture}(20000,14000)

\thicklines

\drawline\fermion[\E\REG](9000,500)[2500]
\drawarrow[\E\ATTIP](\particlebackx,\particlebacky)
\drawline\fermion[\E\REG](\particlebackx,\particlebacky)[2500]
\drawline\photon[\N\REG](\particlebackx,\particlebacky)[17]
\global\advance\pmidx by -2400
\put(\pmidx,\pmidy){$W^{\pm}$}

\drawline\fermion[\NE\REG](\particlefrontx,\particlefronty)[6000]
\drawarrow[\NE\ATTIP](\particlebackx,\particlebacky)
\drawline\fermion[\NE\REG](\particlebackx,\particlebacky)[6000]
\drawline\photon[\E\REG](\particlebackx,\particlebacky)[9]
\global\advance\pmidy by -850
\put(\pmidx,\pmidy){$\gamma$}

\global\Xone=\particlefrontx
\global\Yone=\particlefronty
\global\Xeight=\particlefrontx
\global\Yeight=\particlefronty
\global\advance\Xone by -4243
\global\advance\Yone by 4243
\put(\Xone,\Yone){\circle{3900}}

\drawline\fermion[\N\REG](\Xone,\Yone)[1950]
\drawline\fermion[\S\REG](\Xone,\Yone)[1950]

\global\Xtwo=\Xone
\global\advance\Xtwo by 1000
\drawline\fermion[\N\REG](\Xtwo,\Yone)[1700]
\drawline\fermion[\S\REG](\Xtwo,\Yone)[1700]

\global\Xthree=\Xone
\global\advance\Xthree by -1000
\drawline\fermion[\N\REG](\Xthree,\Yone)[1700]
\drawline\fermion[\S\REG](\Xthree,\Yone)[1700]

\drawline\fermion[\E\REG](\Xone,\Yone)[1950]
\drawline\fermion[\W\REG](\Xone,\Yone)[1950]

\global\Ytwo=\Yone
\global\advance\Ytwo by 1000
\drawline\fermion[\E\REG](\Xone,\Ytwo)[1700]
\drawline\fermion[\W\REG](\Xone,\Ytwo)[1700]

\global\Ythree=\Yone
\global\advance\Ythree by -1000
\drawline\fermion[\E\REG](\Xone,\Ythree)[1700]
\drawline\fermion[\W\REG](\Xone,\Ythree)[1700]

\drawline\fermion[\NW\REG](\Xeight,\Yeight)[2300]
\drawarrow[\NW\ATTIP](\particlebackx,\particlebacky)
\drawline\fermion[\NW\REG](\particlebackx,\particlebacky)[1710]
\global\advance\particlebackx by -2800
\global\advance\particlebacky by 2800
\drawline\fermion[\NW\REG](\particlebackx,\particlebacky)[2200]
\drawarrow[\NW\ATTIP](\particlebackx,\particlebacky)
\drawline\fermion[\NW\REG](\particlebackx,\particlebacky)[1810]

\drawline\fermion[\W\REG](\particlebackx,\particlebacky)[2650]
\drawarrow[\W\ATTIP](\particlebackx,\particlebacky)
\drawline\fermion[\W\REG](\particlebackx,\particlebacky)[2350]

\end{picture}

\caption{A two loop diagram contributiong to the quark-photon vertex.}

\label{fig;quarkem}
\end{figure}
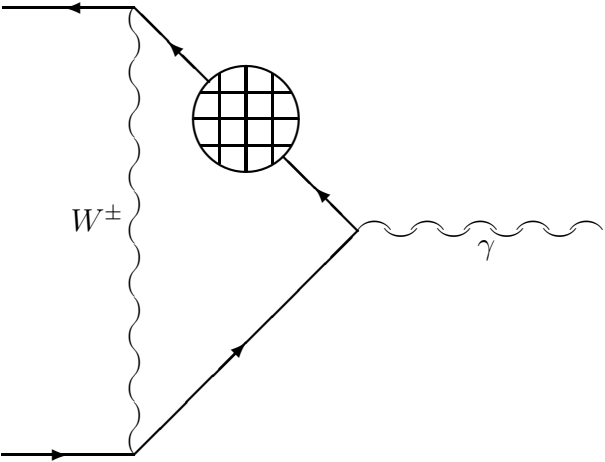

As an illustration of the application of this work to some specific examples,
we will briefly discuss the evaluation of the integral
\be
\mu^{2(4-n)} \int \frac{d^{n}q}{(2 \pi)^{n}} \frac{d^{n}k}{(2 \pi)^{n}}
\frac{1}{k^{2} (k+p)^{2} q^{2} (q-k)^{2}} \label{eq;intex}
\ee
the evaluation of the diagram shown in figure~\ref{fig;bosonse},
and the simplification of the diagram in figure~\ref{fig;quarkem}.
Diagram~\ref{fig;bosonse} is a contribution to the gauge boson self energy
in the presence of massless fermions in some representation R of the gauge
group. The diagram~\ref{fig;quarkem} represents a contribution to the
quark-photon vertex in the Weinberg-Salam model.

Before evaluating the integral (\ref{eq;intex}) we must first declare the list
of four-vector symbols as
\be
{\tt momenta = \{k, p, q, k+p, q+p, k-q, q-k\} \label{eq;mlist}}
\ee
Then we specify the integrand by
\be
{\tt x = \frac{1}{g[k, k]\, g[k+p, k+p]\, g[q, q]\, g[q-k, q-k]}}
\ee
The integral over {\tt q} is performed using inttensor,
\be
{\tt x= inttensor[x, q, k, 1, 1] }
\ee
which results in an expression depending on the integration variable {\tt k.}
The integrand of this expression is proportional to
\be
{\tt \frac{1}{(g[k,k])^{1+ \frac{eps}{2}} \, g[k+p, k+p]}}
\ee
and can also integrated by using {\tt inttensor} as follows.
\be
{\tt x= inttensor[x, k, p, 1+eps/2, 1]}
\ee
The resulting expression is
\be
{\tt \frac{-(2^{2 eps} \, mu^{2 eps}\, pi^{-4\,+\,eps}\, gam[1-eps]\, gam[1-
\frac{eps}{2}]\, gam[\frac{eps}{2}]\, gam[eps])}
{256\, (-1)^{eps}\, g[p,p]^{eps}\, gam[2 -\frac{3 eps}{2}]\, gam[2-eps]\,
gam[1+\frac{eps}{2}]}}
\ee
where {\tt mu} is the introduced mass scale $\mu$,
and this can be Laurent expanded to order $\eps^{-1}$. We obtain
\be
{\tt \frac{-2 \: + \: 2 \, elog \: - \: 5 \, eps}{256\, eps^{2}\, pi^{4}}
+ O( eps)^{0}}
\ee
where
\be
{\tt elog = egam - log \left( \frac{4 \, pi \, mu^{2}}{-g[p,p]} \right)}
\ee
and {\tt egam = 0.577\ldots} is Euler's constant. This gives the result for
two of the four integrals required to determine the pole part
of (\ref{eq;twopt}) and the remaining integrals can be evaluated
similarly. Then to integrate the expression (\ref{eq;twopt}) we first
replace the integrand with an expression that has the same
pole part and then replace the integrals with the Laurent expansions.
The functions we have defined to do this are {\tt samepole} and {\tt poleform}
respectively.

To evaluate the diagram of figure~\ref{fig;bosonse} we declare the
momenta list to be~(\ref{eq;mlist}), and define some initial expressions.
In the notation of our scheme they are the coefficient of the integral,
the numerator and denominator of the integrand. The coefficient is
\be
{\tt coefficient = \frac{i}{2} \, g^{4} \,C[G] \, T[R] \, delta[a, b]}
\ee
where {\tt a} and {\tt b} are gauge group indices, {\tt delta} is the delta
symbol and {\tt C[G]} and
{\tt T[R]} are gauge group factors. The numerator of the integrand has
two factors
\bea
{\tt y = (g[p,a]\, -\, g[k,a]) \, g[u,l] \: +\: (2 g[k,u]\, +\,
g[p,u])\, g[l,a] -} & & \nn \\
{\tt (2 g[p,l] \, + \, g[k,l])\, g[a,u]}\hspace{4.0cm} & &
\eea
and
\be
{\tt z = trace[a.(p+q).v.q.l.(k-q)]}
\ee
and the denominator is
\be
{\tt denominator = g[q,q]\, g[q+p,q+p]\, g[k,k]\, g[k+p,k+p]\, g[k-q,k-q]}
\ee
In the numerator there are two repeated indices. We set
${\tt repeated = \{a,l\},}$ ${\tt other=\{u,v,k,p,q\} }$ and
${\tt x = Expand[y\: z]}$ and
remove the repeated indices from {\tt x} using {\tt contractall,}
\be
{\tt x=contractall[x,repeated,other] }
\ee
All scalar products are then eliminated using {\tt preps} and {\tt prepd}
which make use of identities such as
\be
k.p = \frac{1}{2} ((k+p)^{2} -k^{2} -p^{2}).
\ee
The resulting expression for {\tt x} consists of 69 terms. We then replace
{\tt x} with an expression with the same pole part using {\tt samepole.}
The result is divided by {\tt denominator} and the Laurent expansion is
evaluated by applying {\tt poleform.} Simplifying the result and multiplying by
{\tt coefficient} gives the final result for the diagram
\bea
& & {\tt -( g^{4}\, i\, C[G]\, T[R]\, TR[J]\, delta[a,b] } \nn \\
& & \hspace{0.4cm} {\tt ((-48\: -\: 76 eps\:+\:48 \, elog \: eps)\, g[p,u] \,
g[p,v] \:+ } \nn \\
& & \hspace{0.4cm} {\tt (12\: -59\, eps\: -\: 12 \, elog\:  eps)\, g[p,p]\,
g[u,v]))/(55296 \, eps^2 \, pi^4) }
\eea
where {\tt TR[J]} is the trace of the spin identity matrix.

We have used our scheme to evaluate the gauge boson, fermion and
ghost anomalous dimensions to two loops. We have also calculated the coupling
constant renormalization by evaluating the gauge boson - ghost vertex,
and evaluated the Callan-Symanzik beta function. Our results are in
complete agreement with~\cite{ego79} for these calculations.

Now consider the diagram in figure~\ref{fig;quarkem},
which is a two loop diagram contributing to the quark-photon vertex.
The one loop, flavour-changing quark self energy, with momentum $p$
can be written as
\be
A \, p\!\!\!/ L + B\, p\!\!\!/ R + C\, L + D \, R
\ee
where $A, B, C$ and $D$ are scalars depending upon $p^{2}$ and
the quark masses. After defining
\be
{\tt momenta = \{k,p\} }
\ee
the numerator of the diagram in figure~\ref{fig;quarkem} is proportional,
in our notation, to
\bea
{\tt x} & = & {\tt alpha \, .\, L\, .\, (p+k+M[s])\, .\,
(scalar[A]\, (p\, +\, k)\, .\, L + scalar[B]\,  (p\, +\, k)\, .\, R } \nn \\
& & \hspace{0.4cm} {\tt + scalar[C]\, L + scalar[D]\, R)\, .\, (p+k+M[d])\,
.\, mu \, .\, (p-k+M[d])\, .\,} \nn \\
  &   & \hspace{0.8cm}{\tt alpha. L }
\eea
The masses are implicitly multiplied by the identity matrix.
To simplify this we first apply
\be
{\tt x = movel[x]}
\ee
Then repeated application of the function {\tt con} will do the contractions
over $\ga_{\al}$ and contract the pairs of $k\!\!\!/$'s and $p\!\!\!/$'s.
For example, the following does all possible contractions over
$k\!\!\!/$ and $p\!\!\!/$.
\bea
& & {\tt x\, =\, con[x,p]\, ;\, x\, =\, con[x, k]\, ;\, x\, =\, con[x, p,mu]\,
; } \nn \\
& & \hspace{0.4cm} {\tt \, x\, =\, con[x,k,mu]\, ;\, x\, =\, con[x,p,k]\, ;\,
x\, =\, con[x,k,p]\, ; } \nn \\
& & \hspace{0.8cm} {\tt x\, =\, con[x,p,mu]\, ;\, x\, =\, con[x,k,mu] }
\eea
A further seven applications of {\tt con} are necessary to do all the
contractions over $\ga_{\al}$.
In the resulting expression, repeated use of {\tt comm}
\bea
& & {\tt x\, =\, comm[x,p,k]\, ; \, x\, =\, comm[x,p,mu]\, ;\, x\, =\,
comm[x,k,mu]\, ; } \nn \\
& & \hspace{0.4cm} {\tt x\; =\; comm[x,p,k] }
\eea
followed by collecting with respect to {\tt Dot,}
produces a final expression of four terms, proportional to
$R \ga_{\mu} k\!\!\!/ p\!\!\!/$, $R \ga_{\mu}$, $R k\!\!\!/$ and $R p\!\!\!/$.
The full expression has seventy-six terms when expanded and is too long to
reproduce here.

\section{Conclusion}
We have presented a scheme for representing amputated Feynman diagrams
beyond tree level in terms of Mathematica expressions and have shown how
the algebra involved in simplifying these expressions and evaluating the
integrals can be automated using the symbolic algebra capacity of
Mathematica. This scheme is based upon the specification of a
sufficiently convenient notation for Lorentz tensors and Dirac algebra.
An automatic procedure for the evaluation of pole parts for dimensionally
regularised massless Feynman integrals has been used to evaluate
two loop counterterms in a non-abelian gauge theory with fermions
and for the simplification of amputated amplitudes in the Weinberg-Salam model.
The methods used in this paper can be readily extended to handle
integrals of higher loop order, and integrals involving masses.
Our approach succeeds in eliminating all tedious hand calculations.
The time required to perform calculations automatically is generally small
compared to the time required to prepare input and organise a calculational
sequence.

\section{Acknowledgements}
We would like to thank A. A. Rawlinson and A. C. Kallionatis for their
useful suggestions and assistance.

\section{Appendix}

This appendix contains the usage messages for functions described in this
article.

The following usage messages are for functions which simplify
products of Dirac gamma matrices.
\begin{verse}
{\tt comm::usage = "comm[x, a, b1, b2, \ldots] commutes a to the} \\
{\tt right through b1.b2 \ldots  wherever the pattern a.b1.b2 \ldots } \\
{\tt appears in x (up to 5 b's programmed)."}
\end{verse}
\begin{verse}
{\tt rcomm::usage = "rcomm[x, b1, b2, \ldots, a] commutes a to} \\
{\tt the left through b1.b2 \ldots  wherever the pattern b1.b2 \ldots .a} \\
{\tt appears in x (up  to  5  b's programmed)."}
\end{verse}
\begin{verse}
{\tt con::usage = "con[x, a, b1, b2, \ldots] evalates a.b1.b2 \ldots .a} \\
{\tt by contracting out the a's. Replacement made wherever the} \\
{\tt  pattern appears in x (Up to 6 b's programmed)."}
\end{verse}
\begin{verse}
{\tt trace::usage = "trace[x] evaluates the trace of a linear} \\
{\tt combination of products of gamma matrices that does not} \\
{\tt include gamma\_5."}
\end{verse}
\begin{verse}
{\tt movel::usage = "movel[x] commutes all helicity projection} \\
{\tt operators to the left most position in each term in the} \\
{\tt expression x."}
\end{verse}
\begin{verse}
{\tt mover::usage = "mover[x] commutes all helicity projection} \\
{\tt operators to the right most position in each term in the} \\
{\tt expression x."}
\end{verse}

The following usage messages are for functions which simplify Lorentz tensors.
\begin{verse}
{\tt contract::usage = "contract[x, other, s] will eliminate} \\
{\tt repeated index s from x. A list other, of all other} \\
{\tt symbols appearing in the arguments of g's must be passed} \\
{\tt to the function."}
\end{verse}
\begin{verse}
{\tt contractall::usage = "Given a list of repeated indices,} \\
{\tt repeated, and a list, other, of any indices appearing in} \\
{\tt the arguments of g's that are not in repeated,} \\
{\tt contractall[x, repeated,other] will eliminate all} \\
{\tt of the repeated indices from x."}
\end{verse}
\begin{verse}
{\tt slash::usage = "slash[x, a, b] finds all patterns of the} \\
{\tt form c1.c2. \ldots .ci.a.d1.d2. \ldots .dj*g[a,b] in x and} \\
{\tt replaces them with c1.c2. \ldots .ci.b.d1.d2. \ldots .dj} \\
{\tt where either or both of i and j may be zero."}
\end{verse}
\begin{verse}
{\tt preps::usage = "preps[x, k, p] replaces g[k, p] by} \\
{\tt (g[k+p, k+p] - g[k, k] - g[p, p])/2 everywhere in x."}
\end{verse}
\begin{verse}
{\tt prepd::usage = "prepd[x, k, p] replaces g[k, p] by} \\
{\tt (g[p, p] + g[k, k] -g[k-p, k-p])/2 everywhere in x."}
\end{verse}

The following usage messages are for functions which perform Feynman
integration.
\begin{verse}
{\tt inttensor::usage = "inttensor[x, q, p, a, b, \ldots , j, l]} \\
{\tt integrates tensors (g[q, a] g[q, b] \ldots )/(g[q, q]$\hat{\:}$j
g[q $\pm$} \\
{\tt p, q $\pm$ p]$\hat{\:}$l) of low rank, where the integration variable} \\
{\tt q, external momentum p, powers j and l, and indices} \\
{\tt a,b \ldots are specified in the arguments of inttensor."}
\end{verse}
\begin{verse}
{\tt intgamma::usage = "intgamma[x, q, p, a, b, \ldots , j, l]} \\
{\tt  integrates matrices (q\, g[q, a]\, g[q, b] \ldots )/(g[q, q]$\hat{\:}$j
g[q} \\
{\tt $\pm$ p, q $\pm$ p]$\hat{\:}$l) of low rank, where the integration} \\
{\tt variable q, external momentum p, powers j and l, and} \\
{\tt indices a,b \ldots are specified in the arguments of intgamma."}
\end{verse}
\begin{verse}
{\tt samepole::usage = "samepole[x, p] replaces x, the numerator} \\
{\tt of a Feynman integral for a gauge boson self energy diagram} \\
{\tt with a simpler expression that has the same pole} \\
{\tt part (p is the external momentum)."}
\end{verse}
\begin{verse}
{\tt poleform::usage = "poleform[x, p] evaluates the pole part} \\
{\tt of x, a Feynman integral for a gauge boson self energy} \\
{\tt diagram, after samepole has been applied to the numerator."}
\end{verse}

\pagebreak

\pagebreak

\input FEYNMAN

\begin{figure}
\begin{picture}(20000,10000)

\thicklines

\drawline\gluon[\E\REG](5000,6000)[7]
\drawline\gluon[\NE\REG](\particlebackx,\particlebacky)[5]
\drawline\fermion[\SE\REG](\particlebackx,\particlebacky)[3990]
\drawarrow[\SE\ATTIP](\particlebackx,\particlebacky)
\drawline\fermion[\SE\REG](\particlebackx,\particlebacky)[3990]
\drawline\gluon[\E\REG](\particlebackx,\particlebacky)[7]
\drawline\fermion[\SW\REG](\particlefrontx,\particlefronty)[3990]
\drawarrow[\SW\ATTIP](\particlebackx,\particlebacky)
\drawline\fermion[\SW\REG](\particlebackx,\particlebacky)[3990]
\drawline\gluon[\NW\REG](\particlebackx,\particlebacky)[5]
\drawline\fermion[\N\REG](\particlefrontx,\particlefronty)[5655]
\drawarrow[\N\ATTIP](\particlebackx,\particlebacky)
\drawline\fermion[\N\REG](\particlebackx,\particlebacky)[5655]
\end{picture}

\caption{A two loop diagram contributing to the gauge boson self energy.}

\label{fig;bosonse}
\end{figure}
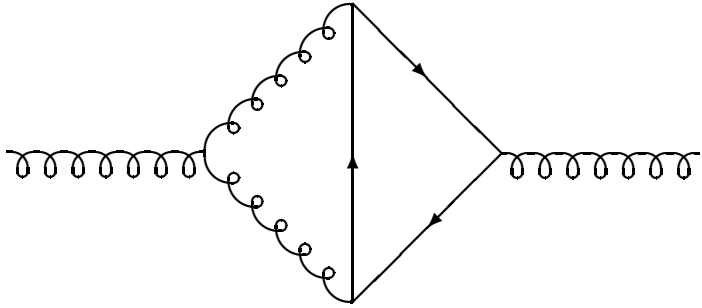

\end{document}